\documentclass[preprint,showpacs]{revtex4}

\usepackage{graphicx}

\begin{document}
\title{THE NON-DISSIPATIVE SPIN-HALL CONDUCTIVITY AND THE IDENTIFICATION OF THE CONSERVED CURRENT}
\author{D.Schmeltzer}
\affiliation{Department of Physics,\\City College of the City University of New York\\
New York,N,Y,10031}
\date{\today}

\begin{abstract}
The two dimensional Rashba Hamiltonian is investigated using the momentum representation.One finds that the $SU(2)$ transformation which  diagonalizes the Hamiltonian gives rise to non commuting Cartesian coordinates for $K=0$ and zero otherwise.This result corresponds to the Aharonov -Bohm phase in the momentum space.

The Spin -Hall conductivity is given by $\frac{e}{4\pi}$ which disagree with the result $\frac{e}{8\pi}$  given in the literature.Using Stokes theorem we find that the Spin -Hall current is carried equally by the up and down electrons on the Fermi surface.

We identify the Magnetic current and find that for an  electric field with a finite Fourier component in the momentum space a non zero Spin -Hall current is obtained.For the electric field which is constant in space ,the orbital  magnetic current cancels the spin current.

In order to measure the Spin-Hall conductance we propose to apply a magnetic field gradient $\Delta H_{2}$ in the $i=2$ direction   and to measure a  Charge- Hall current $\frac{e }{h}(\frac{g}{2})\mu_{B}\Delta H_{2}$ in the $i=1$ direction.

\end{abstract}


\maketitle
\textit{Introduction}.The recent experimental observation of the Spin-Hall effect in a two dimensional electron gas (DEG) with Spin-orbit coupling \cite{wund,aw}, has shown  that a dissipationless spin Hall current might be possible.The origin of this  effect is not clear .The authors in ref. \cite{hir,aw}suggest that the effect is extrinsic while the authors in ref. \cite{wund,simova,olga,halperin,sch} propose that the effect might be intrinsic and dissipationless.
It has been proposed that the dissipationless intrinsic effect might be explained within  the Rashba Hamiltonian \cite{rash,mol,los}. Dissipationless behavior has been also proposed for an other model named the  Luttinger Hamiltonian  \cite{lutt,mura}.

The theory of this effect is far from clear.It is believed that due to rotational invariance of the Rashba Hamiltonian the orbital current cancels the spin current making difficult to measure a spin current\cite{yang}.Other effects such as elastic scattering caused by impurities might be real killers for the Spin-Hall effect in the macroscopic limit \cite{halperin,sch}.

Even the result  for the Spin-Hall current in the absence of disorder is contraversial!The value of the Spin-Hall conductivity  given in the literature is $\frac{e}{8\pi}$  has been obtained within a mean field theory \cite {simova} or by a singular perturbation theory \cite{yang} .

In the first part of this letter we will show that when we  diagonalize  the Rashba Hamiltonian ,non -commuting Cartesian coordinate are induced in the momentum space for zero momentum $\vec{K}=0$ and zero otherwise.This means that the curvature or the fictitious magnetic field is zero for $\vec{K}\neq 0$.This represents the analog of the Bohm -Aharonov flux tube in the momentum space.The result $\frac{e}{8\pi}$ was obtained using a fictitious magnetic field (which originate from the Berry phase ) for electrons in the entire Fermi see. Our derivation is based on the SU(2) transformation which gives a zero curvature (zero fictitious magnetic field) for $\vec{K}\neq 0$.Using Stokes theorem we find that the singular state at $\vec{K}=0$ gives rise to a spin current for the up and down spins on  the Fermi Surface with the  proposed Spin-Hall conductivity  $\frac{e}{4\pi}$!

Due to the difficulty to measure a spin current we propose to measure the Charge-Hall current driven by a magnetic gradient with the conductivity  given by $\sigma_{\bot}=\frac{e g\mu_{B}}{2h}$.

In the second part we introduce the total magnetization and identify the conserved magnetic current.For a slowly varying electric field in space with  a momenta $\vec{Q}\neq 0$ we find a finite spin current which vanishes in the uniform limit $\vec{Q}=0$ (due to the cancellation between the the Spin-Hall current with the orbital magnetic current).

\textit{Non-Commutativity of the Cartesian Coordinate for $\vec{K}=0$ in the Rashba model}.
\begin{equation}
h(\vec{K},\vec{R})=h_{0}(\vec{K})+h_{(ext)}(\vec{R})
\label{equation}
\end{equation}
 $h_{o}(\vec{K})$ is the Rashba Hamiltonian in the momentum space and $h_{(ext)}(\vec{R})$ represents the external potential.The Cartesian coordinates in the momentum representation are given by,$R^{i}=i\partial_{i}\equiv i\frac{d}{d K^{i}}$where $i=1,2.$We will consider two cases :a)The electric field  $E^{(ext)}_{1}$ is applied in the $i=1$ direction and gives rise to the external potential energy ,$h_{(ext)}(\vec{R})\equiv-e E^{(ext)}_{1}\cdot R^{1}$ where $e$ is the electric charge.b)The magnetic field gradient $\Delta H^{(ext,z)}_{2}$ is in the $i=2$ direction with the magnetic potential energy,$h_{(ext)}(\vec{R})\equiv\frac{g\mu_{B}}{2}\sigma^{3}\Delta H^{(ext,z)}_{2}\cdot R^{2}$ \cite{sch} with the magnetic charge $\frac{g\mu_{B}}{2}$ \cite{sch}.
\begin{equation}
h_{o}(\vec{K})=\frac{\hbar^{2}(\vec{K}-{k}_{so}(\vec{\sigma}
\times\hat{e}_{3})
)^{2}}{2m}
\label{Hamiltonian}
\end{equation}
 $k_{so}$ is the spin orbit momentum defined in terms of the effective internal electric field,$\vec{\sigma}$ is the Pauli matrix and $\hat{e}_{3}$ is the unit vector perpendicular to the two dimensional electron gas.

We rotate the spin axes such that the direction of the Pauli matrix $\sigma_{3}$ is  parallel to the vector $\vec{K}$.This rotation is performed with the help of the $SU(2)$ transformation,

 $U(\varphi(\vec{K}),\vartheta=\frac{\pi}{2})=\left(e^{(-i/2)\varphi(\vec{K})\sigma_{3}}\right)\left(e^{(-i/2)\frac{\pi}{2}\sigma_{2}}\right)$. where $\varphi(\vec{K})=\arctan(\frac{- K_{1}}{K_{2}})$ is the azimuth angle in the plane.
This transformation is singular at $\vec{K}=0$.Using this transformation we diagonalize the $h_{0}(\vec{K})$ Hamiltonian.
The transformed Hamiltonian is given by,
 $\hat{h}_{0}(\vec{K})=U^{\dagger}(\varphi(\vec{K}),\vartheta=\frac{\pi}{2})h_{0}(\vec{K})U(\varphi(\vec{K}),\vartheta=\frac{\pi}{2})=\frac{\hbar^{2}(\vec{K})^{2}}{2m}-\sigma_{3}\frac{\hbar^{2}k_{so}|\vec{K}|}{m}\equiv\epsilon(|\vec{K}|,\sigma)$,where $\sigma=\pm$are the two eigenvalues of the Pauli matrix $\sigma^{3}$.
 
Following ref.\cite{sch}we introduce the transformed Cartesian coordinate $\vec{r}$ which replaces the original coordinate $\vec{R}$.Using the unity matrix $I$ we find, 

$I\cdot R^{i}\rightarrow r^{i}=U^{\dagger}(\varphi(\vec{K}),\vartheta=\frac{\pi}{2})I\cdot R^{i}U(\varphi(\vec{K}),\vartheta=\frac{\pi}{2})=I\cdot R^{i}-\frac{\sigma^{1}}{2}\partial_{i}\varphi(\vec{K})$ where  $i=1,2$are the Cartesian coordinates.
As a result we find  the new  commutator.
\begin{equation}
[r^{1},r^{2}]d K^{1}d K^{2}=-i\frac{\sigma^{1}}{2}d(d\varphi(\vec{K})) 
\label{commutator}
\end{equation}
The  commutator in eq.$3$ is expressed with the help of the two form representation $d(d\varphi(\vec{K}))$ which is zero for $\vec{K}\neq 0$ and is singular at $\vec{K}=0$.(The two form is best represented in the complex plane .We introduce the complex coordinates ,$Z=K^{1}+i K^{2}$ and $\overline{Z}=K^{1}-i K^{2}$ and find the complex two form representation,$d(d\varphi(\vec{K}))=\frac{1}{2i}[-\frac{\partial}{\partial Z }(\frac{1}{\overline{Z}})d Z d \overline{Z} +\frac{\partial}{\partial\overline{Z}}(\frac{1}{Z})d \overline{Z}d Z]$.In this complex representation we identify  the two dimensional delta function,  $\delta^{(2)}(\vec{K})= \frac{1}{\pi}\frac{\partial}{\partial\overline{Z}}(\frac{1}{Z})=\frac{1}{\pi}\frac{\partial}{\partial Z }(\frac{1}{\overline{Z}})$)

\textit{The Fermi-Surface Spin-Hall current}.

We will consider first the Spin-current driven by an electric field ,$h_{(ext)}(\vec{R})\equiv-e E^{(ext)}_{1}\cdot R^{1}$.The Spin-Hall current for a  state with momentum  $\vec{K}$ which is  in the Fermi see is represented by by the product between magnetic moment (the Pauli matrix ) and the velocity.

\begin{equation}
J_{2}(\vec{K})=\frac{\hbar}{2}\sigma_{3}\frac{d R^{2}}{d t}=\frac{\hbar}{2}
U^{\dagger}\sigma_{3}U\frac{d r^{2}}{d t}=\frac{\hbar}{2}(-\sigma^{1})\frac{d r^{2}}{d t}
\label {state}
\end{equation}
In the second part of  equation $4$ we have used the transformed magnetic moment ($\sigma^{3}\rightarrow U^{\dagger}\sigma_{3}U$)and the transformed velocity ($\frac{d R^{2}}{d t}\rightarrow\frac{d r^{2}}{d t}$).
The velocity in the $i=2$ direction is obtained from the Heisenberg equation of motion.
\begin{equation}
i\hbar\frac{d r^{2}}{d t}=[r^{2},\hat{h}_{o}(\vec{K})]-e E^{(ext)}_{1}[r^{2},r^{1} ] 
\label{Heisenberg}
\end{equation}
The current driven by the electric field is obtained by substituting in the current equation $4$the velocity given  by the last term in eq.5.In order to compute the current we have to consider the current contribution $J_{2}(\vec{K})$from the full Fermi see.This current is obtained by using the $Fermi$ -$Dirac$ occupation function at $T=0$, $\theta[\epsilon(|\vec{K}|,\sigma)-E_{F}]$.The total current will be given by  the two dimensional integral over the Fermi see .This corresponds to the  current density $J_{2}(\vec{K})$ multiplied by the   $Fermi$-$Dirac$ function $\theta[\epsilon(|\vec{K}|,\sigma)-E_{F}]$.

\begin{equation}
J_{2}=\frac{1}{(2\pi)^{2}}\sum_{\sigma}\int\int J_{2}(\vec{K})\theta[\epsilon(|\vec{K}|,\sigma)-E_{F}]d K^{1}d K^{2}=\frac{e}{2}E^{(ext)}_{1}\frac{1}{2}\sum_{\sigma}\oint \frac{d\varphi}{(2\pi)^{2}}=\frac{e}{4\pi}E^{(ext)}_{1}
\label{spin current}
\end {equation}

The second part of equation $6$ is obtained with the help of the $Stokes$ theorem .The theorem allows us to replace the two dimensional integral over the full Fermi see with  an integral over the Fermi surface (the two Fermi surfaces ,for spin up and spin down)only.The value of the Fermi surface line integral is obtained with the help of the two dimensional delta function  ($\delta^{(2)}(\vec{K})= \frac{1}{\pi}\frac{\partial}{\partial\overline{Z}}(\frac{1}{Z})=\frac{1}{\pi}\frac{\partial}{\partial Z }(\frac{1}{\overline{Z}})$).
As a result we find that the $Spin$-$Hall$ current is given  by the $Fermi$-$Surface$ contribution only.We believe that the value of the conductivity $\frac{e}{4\pi}$ obtained here is a direct consequence of the Fermi-Surface contribution.

\textit{The Fermi-Surface Charge -Hall current}.

The second case which  we consider corresponds  to a  magnetic field gradient in the $i=2$ direction which induces  a $Charge$-$Hall$ current for the  state $\vec{K}$ in the $i=1$ direction .This charge current is given by the product charge-velocity ,
 $J_{1}(\vec{K})=-e\frac{d r^{1}}{d t}$.

The velocity in the $i=1$ direction is obtained from the Heisenberg equation of motion,
\begin{equation}
i\hbar\frac{d r^{1}}{d t}=[r^{1},\hat{h}_{o}(\vec{K})]+\frac{g\mu_{B}}{2}U^{\dagger}\sigma^{3}U\Delta H^{(ext,z)}_{2}[r^{1},r^{2} ] 
\label{Heisenbergm}
\end{equation}
Following the same steps as used for the $Spin$-$Hall$ current  given in equations $4-6$ we find from eq.$7$ the $Charge$-$Hall$ current, 
$J_{1}=\frac{e}{h}\frac{g}{2}\mu_{B}\Delta H^{(ext,z)}_{2}$

Due to the difficulty to measure a spin current we propose \cite{sch}to measure the Spin-Hall conductivity by measuring an electric current in the $i=1$ direction driven by a magnetic gradient in the $i=2$ direction.

\textit{The Conserved Magnetic Current}.

One of the difficulties of  the Spin-Hall theory is the proper identification of the spin current.In order to accomplish
this goal we will identify the total magnetization of the system as the conserved quantity.The total magnetization in two dimensions has two parts, the macroscopic orbital magnetization and the $z$ component of the spin density.Using the Hamiltonian of the system and the Heisenberg equation of motion we will obtain the continuity equation for the magnetization. This  will allow us to identify the conserved spin current.

We will work in the second quantized form.We will introduce the two component Spinor operator $\Psi(\vec{K})$. The 
magnetization density which is a function of the wave vector $\vec{q}$ will be expressed in the second quantized form using the two component Spinor $\Psi(\vec{K})$.
The magnetization  density $M(\vec{q})$ contains two parts, the spin density given by the Pauli  $\sigma^{3}$component and the uniform orbital magnetization ,$\vec{R}\times\vec{K}$.
\begin {equation}
M(\vec{q})=\hbar\int\int\frac{d^{2}K}{(2\pi)^{2}}\Psi\dagger (\vec{K})[\frac{\sigma^{3}}{2}+(\vec{R}\times\vec{K})\delta_{(\vec{q},0)}]e^{i\vec{q}\cdot\vec{R}}\Psi(\vec{K})
\label{magnetization}
\end{equation}
The operator $e^{i\vec{q}\cdot\vec{R}}$ acts to shift the momenta $\vec{K}$to $\vec{K}-\vec{q}$.

The Hamiltonian which we will use to identify the current operator is the second quantized form of the Hamiltonian given in e q.$1$ ,
\begin{equation}
H=\int\int\frac{d^{2}K}{(2\pi)^{2}}\Psi\dagger(\vec{K})[h_{0}(\vec{K})+\int\frac{d Q_{1}}{2\pi}h^{(ext)}(Q_{1})e^{i Q_{1}\cdot R^{1}}]\Psi(\vec{K})
\label{second hamiltonian}
\end{equation}
The Hamiltonian given in eq.$9$ is driven by the external potential $h^{(ext)}(Q_{1})$ expressed in terms of the one component Fourier  electric field $E^{(ext)}_{1}(Q_{1})$of momenta $Q_{1}$ in the $i=1$ direction.
\begin{equation}
h^{(ext)}(Q_{1})=-e\frac{E^{(ext)}_{1}(Q_{1})}{i Q_{1}}\delta(Q_{1}-q_{1})
\label{external}
\end{equation}
Next we use the Heisenberg equation of motion for the Spinors $\Psi^{\dagger}(\vec{K},t)$ and $\Psi(\vec{K},t)$to obtain the continuity equation  .We find,

\begin{eqnarray}\nonumber
\partial_{t}M(\vec{q})&=& -i\int\int\frac{d^{2}K}{(2\pi)^{2}}\Psi\dagger (\vec{K})\left([\frac{\sigma^{3}}{2}+(\vec{R}\times\vec{K}),h_{o}(\vec{K})]\delta_{(\vec{q},0)} \right. \nonumber \\&+&\int\frac{d Q_{1}}{2\pi}h^{(ext)}(Q_{1})[e^{i Q_{1}\cdot R^{1}},(\frac{\sigma^{3}}{2}+(\vec{R}\times\vec{K}))\delta_{(\vec{q},0)}e^{i \vec{q}\cdot \vec{R}}]\nonumber\\&+&[\frac{\sigma^{3}}{2}(e^{i\vec{q}\cdot\vec{R}}-\delta_{(\vec{q},0)}),h_{o}(\vec{K})]\nonumber\\&+& \left.
(1-\delta_{(\vec{q},0)})\int\frac{d Q_{1}}{2\pi}h^{(ext)}(Q_{1})[e^{i Q_{1}\cdot R^{1}},(\frac{\sigma^{3}}{2})e^{i \vec{q}\cdot \vec{R}}] \right)
\Psi(\vec{K})
\end{eqnarray}
We observe that the first commutator in e q.$11$ is zero since the total momentum is conserved and therefore commutes with the rotational invariant  Hamiltonian $h_{o}(\vec{K})$.
In eq.$11$ we have used the delta function $\delta_{(\vec{q},0)}$ to separate the uniform part  ($\vec{q}=0$)from the non-uniform one ($\vec{q}\neq0$).

In the next step we perform the SU(2) unitary transformation .As a result the Hamiltonian $h_{o}(\vec{K})$is replaced by the transformed Hamiltonian which is diagonal in the spin and momentum space,
$\hat{h}_{0}(\vec{K})=\frac{\hbar^{2}(\vec{K})^{2}}{2m}-\sigma_{3}\frac{\hbar^{2}k_{so}|\vec{K}|}{m}\equiv\epsilon(|\vec{K}|,\sigma)$.
The Cartesian coordinate are replaced ,$\vec{R}\rightarrow\vec{r}$ with the non commuting properties,$[r^{1},r^{2}]d K^{1}d K^{2}=-i\frac{\sigma^{1}}{2}d(d\varphi(\vec{K}))$, in agreement with eq.$3$. The many -body ground state is replaced with the help of the SU(2) transformation.We obtain the 
the Fermi see state $|F.S.>=|K_{F}^{\uparrow},K_{F}^{\downarrow}>$ .This state represents the filled Fermi see   up to the two Fermi momentum,  $K_{F}^{\uparrow}$and $K_{F}^{\downarrow}$.The  Spinor which acts on the state $|F.S.>=|K_{F}^{\uparrow},K_{F}^{\downarrow}>$ are $\hat{\Psi}\dagger(\vec{K})$and $\hat{\Psi}(\vec{K})$  and replaces the original Spinor $\Psi\dagger(\vec{K})$and $\Psi(\vec{K})$.

The second commutator in eq.$11$ is  zero .This follow from the fact that the transformed total angular momentum is given by ,$U^{\dagger}[\frac{\sigma^{3}}{2}+(\vec{R}\times\vec{K})]U=(\vec{R}\times\vec{K})$. This result shows that the $SU(2)$ transformation  induces a shift on the Pauli matrix $\sigma^{3}$ which cancels the changes induced on   the Cartesian coordinate $\vec{R}$ (the difference between  the coordinate $\vec{R}$ and the transformed coordinate $\vec{r}$ is given by $U^{\dagger}\frac{\sigma^{3}}{2}U$).As a result  the expectation value with respect to the state with a zero orbital angular momentum
  $|F.S.>=|K_{F}^{\uparrow},K_{F}^{\downarrow}>$ is zero.Therefore a uniform electric field with $Q_{1}=0$ momentum
gives rise to a spin current which is canceled by the orbital part. 

The only current which is not zero occurs for  a non uniform electric fields with a non zero momentum.

In order to find the continuity equation we evaluate the third and fourth  commutators given in  eq.$11$ which are finite.
The third commutator gives us the regular velocity contribution (see the first term in eq.$12$and eq.$13$).
The fourth commutator in eq.$11$ is determined by the exponential of the coordinates commutator,
which is given by,$[e^{i Q_{1}\cdot r^{1}},e^{i q_{1}\cdot r^{1}}e^{i q_{2}\cdot r^{2}}]=e^{i (Q_{1}\cdot r^{1}+q_{1}\cdot r^{1})}e^{i q_{2}\cdot r^{2}}[e^{q_{2}\cdot Q_{1}[r^{1},r^{2}]}-1]$.We will use this result in order to determine the electric field contribution given by the external potential in  eq.$10$ We will substitute in eq.$11$ the space dependent electric $E^{(ext)}_{1}(Q_{1})$  given in eq.$10$ which satisfies $Q_{1}=q_{1}\neq 0$. 
Using the third and fourth commutator given by eq.$11$ we  find in the long wave limit the $continuity$-$equation$ , $\partial_{t}M(\vec{q})+i\vec{q}\cdot\vec{J}(\vec{q})=0$.
From the continuity equation we identify the currents for a finite momentum $\vec{q}$ in the $i=1$and $i=2$ directions .

\begin{equation}
J_{1}(\vec{q})=\int\int\frac{d^{2}K}{(2\pi)^{2}}\hat\Psi^{\dagger}(\vec{K})\left(-\frac{\sigma^{1}}{2i}[r^{1},\hat{h}_{0}(\vec{K})]e^{i\vec{q}\cdot\vec{R}}\right)\hat\Psi(\vec{K})
\label{curentone}
\end{equation}

\begin{eqnarray}
J_{2}(\vec{q})&=& \int\int\frac{d^{2}K}{(2\pi)^{2}} \hat\Psi^\dagger(\vec{K}) \left(-\frac{\sigma^{1}}{2i}[r^{2},\hat{h}_{0}(\vec{K})]e^{i\vec{q}\cdot\vec{R}} \right) \hat\Psi(\vec{K}) \nonumber\\&+&  e E^{(ext)}_{1}(Q_{1}=q_{1})\int\int\frac{d^{2}K}{(2\pi)^{2}}\hat\Psi^\dagger(\vec{K})\left(-\frac{\sigma^{1}}{2i}[r^{1},r^{2}] \right) \hat\Psi(\vec{K})
\end{eqnarray}
The first terms in eqs.$12,13$ represent the transformed spin current in the absence of the electric field. In particular the first term in eq.$13$ has the same transformed form as given in eq.$4$.The second term in eq.$13$ represents the $Spin$-$Hall$ current driven  by the non-uniform electric field and is similar to the result given in eqs.$5-6$ (the explicit form is the  same as given by the last term in eq.$5$ which is substituted in eq.$4$ and used in eq.$6$ to perform the $Stoke$ integral).

In order to compute the current we have to compute the expectation value with respect the ground state .To lowest order we compute the expectation value with respect $|F.S.>=|K_{F}^{\uparrow},K_{F}^{\downarrow}>$.As a result we find that  the $Spin$- $Current$ in the $i=1$ direction is zero, $<F.S.|J_{1}(\vec{q})|F.S.>=0$ .(The next order term will given the dissipative current in the $i=1$ direction driven by the electric field $E^{(ext)}_{1}(Q_{1}=q_{1})$.)

The $non$ $dissipative$ $Spin$-$Hall$ $current$ in the $i=2$ direction is non zero for $\vec{q}\neq 0$: 
$<F.S.|J_{2}(Q_{1}=q_{1},q_{2}=0)|F.S.>=\frac{e}{4\pi}E^{(ext)}_{1}(Q_{1}=q_{1})$ ,this result is similar to the result given for a uniform electric field obtained in eq.$6$.  

\textit{The effect of disorder-Non-Dissipative case}.
The effect of disorder is investigated by adding to the Hamiltonian in eq.$1$ a random potential.As a result the ground state wave function is changed from  $|F.S.>$ to the new ground state $|F.S.>>$ ,\cite{sch}.The current is obtained by computing the new expectation function $<<F.S.|J_{2}(Q_{1}=q_{1},q_{2}=0)|F.S.>>$.This expectation function is expressed in terms of the    the averaged single particle time ordered Green's functions ,see ref.\cite{don},  pages  103-109 and ref.\cite{sch}.(Here we will not consider the dissipative part. The dissipative part is obtained within the linear response theory see eq.$5.5.12$ in \cite {don}) ).The effect of disorder is in particular strong for the state $\vec{K}=0$,since for this state the single particle wavelength is larger than the elastic mean free path  suggesting that the momentum representation might not be valued. (This situation might be solved for finite systems where the elastic mean free path is comparable with the size of the system.)Using the averaged single particle time ordered Green's functions in the momentum -and frequency space,  $\overline{G_{\uparrow,\uparrow}(\vec{K},\vec{K};\omega)}$,$\overline{G_{\downarrow,\downarrow}(\vec{K},\vec{K};\omega)}$ ( see ref.     \cite{sch}) allows us to compute  the Spin-Hall current. 

$J_{2}=E^{(ext)}_{1}\int\frac{d\omega}{4\pi i}\int d^{2}K\delta^{2}(K)[\overline{G_{\uparrow,\uparrow}(\vec{K},\vec{K};\omega)} +\overline{G_{\downarrow,\downarrow}(\vec{K},\vec{K};\omega)}]\leq\frac{e}{4\pi}E^{(ext)}_{1}$

The Spin-Hall current is determined by the residue of the single particle Green's function . The Fermi-surface current is finite only if all the states below the Fermi-surface have a finite residue. This might be not satisfied for the state $\vec{K}=0 $ which is sensitive to disorder.As a result in the macroscopic limit the current might be zero.

To conclude, an exact value for the $Spin$-$Hall$ conductivity given by $\frac{e}{4\pi}$ and carried by the electrons on the $Fermi$-$Surface$ has been found ,the $conserved$ $spin$ $current$ for a finite momentum has been identified and a new way  to measure the $Spin$-$Hall$ conductivity based on a magnetic field gradient which induces a $Charge$-$Hall$ current has been proposed.


\begin{figure}
     \includegraphics[width = 4 in]{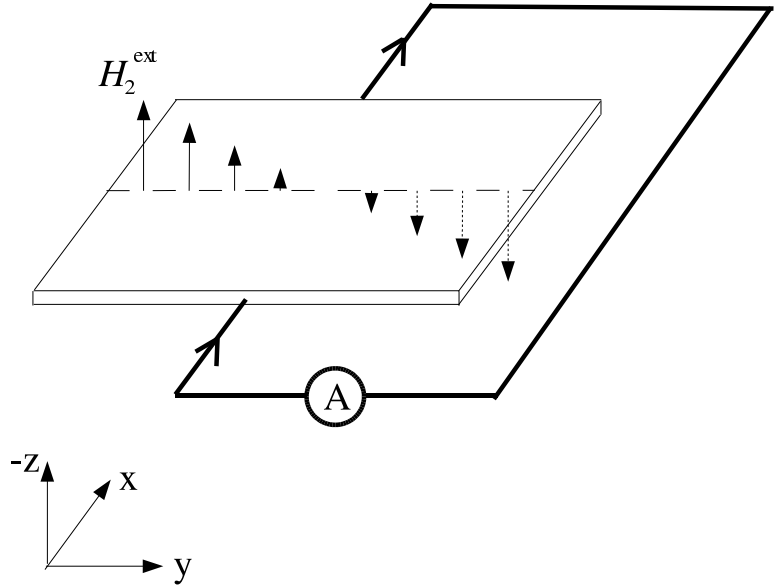}
     \caption{shows the Hall current driven by a magnetic gradient .}
\end{figure}


\begin{thebibliography}{99}

\bibitem{wund}J.Wunderlich, B.Kastner, J.Simova, and T.Jungwirth, Phys.Rev.Lett.\textbf{94}, 047204 (2005).
\bibitem{aw}Y.K.Kato, R.C.Myers, A.C.Gossard and D.D.Awschalom, Science \textbf{306},1910(2004).
\bibitem{hir}J.E.Hirsch, Phys.rev.Lett.\textbf{83}, 1834 (1999).
\bibitem{simova}J.Simova, D.Culcer, Q.Niu, N.A.Sinitsyn, T.Jungwirth and A.H.Macdonald, Phys.Rev.Lett.\textbf{92}, 126603(2004).
\bibitem{olga} O.V.Dimitrova, cond-mat/0405339.
\bibitem{halperin} E.G.Mishenko, A.V.Shytov. and B.I.Halperin, cond-mat/0406730.
\bibitem{sch}D.Schmeltzer,cond-mat/0406565.
\bibitem{mura}S.Muramaki, N.Nagaosa, S-C.Zhang, cond-mat/040660.
\bibitem{rash}Yu.A.Bychov and E.I.Rashba, J.Phys.C\textbf{17}, 6039(1984).
\bibitem{mol}L.W.Molenkamp, G.Shmidt and G.E.Bauer, cond-mat/014109.L.W.Molenkamp, Phys.rev.B\textbf{67}, 033104(2003) and cond-mat/0420442.
\bibitem{los}J.Schliemann and D.Loss, cond-mat/0306523.
\bibitem{lutt}J.M.Luttinger, Phys.Rev.\textbf{102}, 1030 (1956).
\bibitem{yang}S.Zhang and Z.Yang, Phys.Rev.Lett.\textbf{94}, 066602(2005).
\bibitem{don}S.Doniach and E.H.Sondheimer, "Green's functions for solid State Physics", pages 111-112. World Scientific Publishing, second edition ,1998.
\end{thebibliography}
\end{document}